\documentclass[12pt]{article}
\usepackage{wrapft}
\usepackage{bm}

\begin{document}

\title{Soft Supersymmetry Breaking Masses and $\mu$ Parameter from Dynamical Rearrangement
of Exotic $U(1)$ Symmetries}

\author{
Yoshiharu \textsc{Kawamura}\footnote{E-mail: haru@azusa.shinshu-u.ac.jp} 
and Takashi \textsc{Miura}\footnote{E-mail: s09t302@shinshu-u.ac.jp}\\
{\it Department of Physics, Shinshu University, }\\
{\it Matsumoto 390-8621, Japan}
}

\date{
June 1, 2011}

\maketitle
\begin{abstract}
We propose a mechanism that the soft supersymmetry breaking masses
and $\mu$ parameter can be induced
from the dynamical rearrangement of local $U(1)$ symmetries
in a five-dimensional model.
It offers to a solution of $\mu$ problem 
if there is a large hierarchy among the relevant $U(1)$ charge of Higgsinos 
and that of hidden fields which stabilize the extra-dimensional component of $U(1)$ gauge boson.
\end{abstract}

\section{Introduction}

The gauge hierarchy problem (naturalness problem) can be solved by the supersymmetry (SUSY) 
because SUSY stabilizes the weak scale against radiative corrections from a higher-energy physics.\cite{V}
It is also expected that SUSY can play an essential role in physics at the Planck scale
because string theory is a powerful candidate as an unified theory including quantum gravity
and its consistency is deeply related to the worldsheet modular invariance and SUSY.

If nature takes advantage of SUSY, 
SUSY should be broken with soft SUSY breaking parameters of $O(1)$ TeV
because of naturalness and yet-to-be-discovered superpartners.
Then, the origin of soft SUSY breaking terms and $\mu$ term is one of the biggest problem 
beyond the standard model (SM) based on SUSY.\cite{K&N}
The minimal SUSY extension of SM is so-called the MSSM.
It is usually expected that 
a high-energy physics is described by a quantum field theory (QFT) respecting SUSY, 
the SUSY is spontaneously broken in some hidden sector,
and soft SUSY breaking terms are induced in our visible sector by the mediation of some messengers.\cite{N}

Based on the bottom-up approach and the brane world scenario, 
we explore a possibility that SUSY is not completely realized in our starting high-energy QFT.
{\it A lesson from the brane world scenario is that symmetries are not necessarily realized uniformly
over the space-time including extra dimensions.}\cite{M&P,K1} 
We present the exotic scenario that 
SUSY is assumed to be explicitly broken, at some high energy scale, in the presence of extra gauge symmetries
which interact fields and their superpartners differently in the bulk, 
but our brane respects the $N=1$ SUSY.
The physics on our brane is described as the MSSM with
the soft SUSY breaking terms and $\mu$ term, which are originated from  
the dynamical rearrangement of broken gauge symmetries.

The dynamical rearrangement is a part of Hosotani mechanism.\cite{H1,H2,HHHK}
The physical symmetry and spectra are obtained after the determination of vacuum state via the mechanism.
An excellent feature is that the physics is mostly dictated by the particle contents of the theory
including the assignment of gauge quantum numbers.
It is interesting to pursue what type of particle contents and structure of extra dimensions
are appropriate to derive phenomenologically 
suitable soft SUSY breaking terms and $\mu$ term
by setting the gauge invariance above SUSY.

In this paper, we propose a mechanism that the soft SUSY breaking masses and $\mu$ parameter can be induced 
from the dynamical rearrangement of local $U(1)$ symmetries on the basis of the above scenario. 
As a bonus, it offers to a solution of $\mu$ problem if there is a large hierarchy among the relevant $U(1)$ charge of Higgsinos 
and that of SM singlet fields which stabilize the extra-dimensional component of $U(1)$ gauge boson. 
In the next section, we elaborate our scenario after explaining our background, and give our setup on the model building.
In section 3, we give a candidate for realizing our scenario.
In the last section, we present conclusions and a discussion.

\section{Our scenario and setup}

\subsection{Background}

First it would be meaningful to explain the relevant preceding study for the origin of soft SUSY breaking terms from extra dimensions.

Scherk and Schwarz proposed the mechanism that SUSY breaking terms originate from the different boundary conditions (BCs)
between fields and those superpartners.\cite{S&S1,S&S2} 
This is so-called Scherk-Schwarz mechanism.
It is applied to the SUSY SM on the five-dimensional space-time and 
the MSSM is derived on the brane.\cite{P&Q,BH&N1,BH&N2}
The magnitude of soft SUSY breaking masses ($m_{\tiny{\mbox{SUSY}}}$) and $\mu$ parameter is $O(\alpha_l/R)$
where $\alpha_l$ are twisted phases relating BCs.
We refer to them as Scherk-Schwarz phases.
Tiny phases of $\alpha_l = O(R/\mbox{TeV}^{-1})$ are necessary (in the unit of TeV$^{-1}$ for the radius of extra dimension $R$)
in order to obtain $m_{\tiny{\mbox{SUSY}}}$ and $\mu$ of  $O(1)$TeV.
Furthermore the non-abelian symmetry such as $SU(2)_R$ and $SU(2)_H$ can be crucial 
for the mechanism
to work because the Scherk-Schwarz phases are attached to the doublets of $SU(2)_R$ and $SU(2)_H$
to generate $m_{\tiny{\mbox{SUSY}}}$ and $\mu$, respectively.
Here $SU(2)_R$ is the $R$ symmetry and $SU(2)_H$ is the symmetry between two sets of hypermultiplets.

According to the Hosotani mechanism, 
the system with the non-vanishing vacuum expectation values (VEVs) of Wilson line phases and ordinary periodic BCs 
is gauge equivalent to that with the vanishing VEVs of Wilson line phases and twisted BCs 
via the dynamical rearrangement of relevant gauge symmetries.
Based on this observation, we hit on the idea whether soft SUSY breaking parameters and $\mu$ parameter
stem from the Hosotani mechanism.
There are, however, two great barriers.
One is that a non-vanishing effective potential cannot be induced from any global SUSY theory
without SUSY breaking terms.
The other is that $SU(2)_R$ cannot be a local symmetry in any global SUSY theory
because $SU(2)_R$ does not commute with SUSY.

These difficulties can be avoid by extending the framework of theory.
In fact, Gersdorff and Quir\'os pointed out that the Scherk-Schwarz mechanism
can be interpreted as the Hosotani mechanism relating to the local $SU(2)_R$ symmetry 
in the supergravity (SUGRA) on the five-dimensional space-time 
including the orbifold $S^1/Z_2$ as the extra space.\cite{G&Q}
Furthermore Gersdorff, Quir\'os and Riotto calculated the effective potential and found that
the Scherk-Schwarz phase is fixed to 0 or $\pi$ at the one loop level.\cite{GQ&R}
Then $m_{\tiny{\mbox{SUSY}}}$ of $O(1/R)$ can be obtained because the normalization of $SU(2)_R$ charges is fixed
from the group theoretical reason.
This result is rather robust in the absence of other SUSY breaking sources 
because the effective potential depends on the particle content in the bulk.

In this way, we have arrived the no go theorem that 
{\it soft SUSY breaking parameters and $\mu$ parameter of $O(1)$TeV cannot be obtained via the Hosotani mechanism
from any SUSY QFT without SUSY breaking sources and with flat small extra dimensions.}

\subsection{Scenario}

It is necessary to relax some of assumptions to escape the above no go theorem.
We adopt the assumption that SUSY in starting QFT is partially broken by some symmetries.
The basic ingredients of our scenario are as follow.
\begin{itemize}
\item The space-time is made from a product of four-dimensional Minkowski space $M^4$ and the extra space. 
Our four-dimensional world is a brane or boundary in the bulk.

\item Gauge bosons live in the bulk and the relevant gauge group is $G_{\rm SM} \times G'$.
Here $G_{\rm SM}$ is the SM gauge group $G_{\rm SM} = SU(3)_C \times SU(2)_L \times U(1)_Y$
and $G'$ is an exotic gauge group.

\item A same number of bosonic fields and fermionic ones exits, e.g., as a remnant of SUSY at a higher energy scale
beyond our starting QFT.
The corresponding partners have a same quantum number under $G_{\rm SM}$,
{\it but a different quantum number under $G'$}.\footnote{
We do not specify the origin of exotic gauge symmetries.
As a conjecture, bulk fields with $G'$ quantum numbers might be solitonic objects originating from unknown non-perturbative dynamics
on the formation of space-time structure in a more fundamental theory.
Or they might be survivers from SUSY multiplets after decoupling some partners.
}
Hence {\it the SUSY is manifestly broken in the bulk at a turn of the switch of $G'$.}

\item The $G'$ is broken down to its subgroup $H'$ on our brane
by suitable BCs relating the extra dimension, which respect the SUSY.
Hence {\it the $N=1$ SUSY can be an unbroken symmetry on our brane} if all fields are singlets
(or fields and their would-be superpartners have a common quantum number) under $H'$.

\item The MSSM fields come from zero modes of bulk fields and
our brane is described by a SUSY theory with a symmetry
such as $R$ parity, which prevents from decaying a proton rapidly.

\item Non-singlet fields under the broken symmetries $G'/H'$ can obtain masses 
from the dynamical rearrangement of $G'/H'$ via the Hosotani mechanism.
The mass terms become SUSY breaking ones 
if the $G'/H'$ quantum numbers for fields are different from those of their superpartners.

\item The masses are proportional to the charges of $G'/H'$ and the VEVs of
extra-dimensional components of broken gauge bosons (Wilson line phases).
Hence the $m_{\tiny{\mbox{SUSY}}}$ and $\mu$ of  $O(1)$TeV can be obtained
in the case that the charges of $G'/H'$ and/or the VEVs of Wilson line phases are tiny enough.
\end{itemize}

\subsection{Setup}

We consider an extension of MSSM on the five-dimensional space-time $M^4 \times S^1/Z_2$.
The coordinates $x^{M}$ $(M =  0,1,2,3,5)$ are separated into 
the uncompactified four-dimensional ones $x^{\mu}$ $(\mu =  0,1,2,3)$ (or $x$)
and the compactified one $x^5$ (or $y$).
The $S^1/Z_2$ is obtained by dividing the circle $S^1$ (with the identification $y \sim y + 2\pi R$) 
by the $Z_2$ transformation $y \to -y$, so that the point $y$ is identified with $-y$.
Then the $S^1/Z_2$ is regarded as an interval with length $\pi R$, with $R$ being the $S^1$ radius.
Both end points $y=0$ and $\pi R$ are fixed points under the $Z_2$ transformation.
We regard the four-dimensional hypersurface on $y=0$ as our visible world.
 
We line up a $U(1)$ gauge symmetry as the candidate of broken exotic symmetry
because any tiny $U(1)$ charge is allowed theoretically as far as a quantization condition is not imposed.
Here we have a problem that the fifth component of extra $U(1)$ gauge boson ($A_{5}$)
has usually $Z_2$ odd parities and it cannot play the role of Wilson line phase.
There are two ways to make $Z_2$ parities of $A_5$ even.
One is that we impose the conjugate BCs on fields.\cite{HK&O}
But, this type of BCs are not suitable for the SM gauge bosons
and matter fields because zero modes of the SM gauge bosons are projected out
and zero modes of matter fields possess only real components.
The other is that we use a variant of the diagonal embedding proposed in Ref.~\cite{KTY}.
As shown later, we can impose appropriate BCs on both SM gauge bosons
and matter fields, and the non-abelian structure such as $SU(2)$ can be build
by making a difference between the eigenstates of $U(1)$ gauge symmetry and those of BCs.
Hence we adopt the latter one.

We consider $U(1)^{(-)}$ as $G'$
and impose the following BCs on the gauge boson $A_{M}^{(-)}$,
\begin{eqnarray}
\hspace{-0.7cm} &~& A_M^{(-)}(x,y+2\pi R)=A_M^{(-)}(x,y)~,~~
\label{Apm-BC1}\\
\hspace{-0.7cm} &~& A_{\mu}^{(-)}(x,-y)=- A_{\mu}^{(-)}(x,y)~,~~ A_{5}^{(-)}(x,-y)= A_{5}^{(-)}(x,y)~,
\label{Apm-BC2}\\
\hspace{-0.7cm} &~& A_{\mu}^{(-)}(x, 2\pi R-y)=- A_{\mu}^{(-)}(x,y)~,~~ A_{5}^{(-)}(x,2\pi R-y)= A_{5}^{(-)}(x,y)~.
\label{Apm-BC3}
\end{eqnarray}
{}From (\ref{Apm-BC1}) -- (\ref{Apm-BC3}),
we find that the $U(1)^{(-)}$ is broken down on our brane
because the massless mode do not appear in $A_{\mu}^{(-)}$ but $A_{5}^{(-)}$.
In this case, $H'$ is nothing.
The massless mode of $A_{5}^{(-)}$ is a dynamical field which will play a central role as the Wilson line phase
in the dynamical rearrangement.

Next we consider a pair of fermions $\Psi_k$ and a pair of complex scalar bosons $\Phi_k$
whose Lagrangian density is given by
\begin{eqnarray}
\mathcal{L} = \sum_{k=1, 2} i \overline{\Psi}_k \Gamma^M D_M \Psi_k 
 + \sum_{k=1,2} |D_M \Phi_k|^2~,
\label{Psi-L}
\end{eqnarray}
where $\Gamma^M$ is a five-dimensional gamma matrices and $D_M$ is the covariant derivative defined by
\begin{eqnarray}
 D_M \equiv \partial_M + i g_- Q_- A_M^{(-)} + i \sum_{a=1}^3 g_a A_M^{a}~.
\label{D}
\end{eqnarray}
Here $g_-$ and $Q_-$ are the gauge coupling and the charge of $U(1)^{(-)}$, and 
$g_a$ and $A_M^{a}$ are the gauge couplings and the gauge bosons of $G_{\rm SM}$.
We need a pair of fields whose $U(1)^{(-)}$ charge has an opposite value,
in order to realize the invariance under the $Z_2$ transformation.

Now we study BCs and one-loop effective potential of $A_{5}^{(-)}$ for fermions and complex scalar fields in order.

\subsubsection{Fermions}

We consider a pair of Dirac fermions $\Psi_1$ and $\Psi_2$ whose $U(1)^{(-)}$ charge
is $q_-$ and $-q_-$, respectively.
The following BCs are compatible with that the Lagrangian density (\ref{Psi-L}) should be a single-valued function
on $M^4 \times S^1/Z_2$,
\begin{eqnarray}
\hspace{-0.7cm} &~& \Psi_k(x,y+2\pi R)=\eta_{f0}\Psi_k(x,y)~~~(k=1, 2)~,
\label{Psi-BC1}\\ 
\hspace{-0.7cm} &~& \Psi_1(x,-y)=\eta_{f1}\gamma_5\Psi_2(x,y)~,~~
\Psi_1(x,2\pi R-y)=\eta_{f2}\gamma_5\Psi_2(x,y)~,
\label{Psi-BC2}
\end{eqnarray}
where $\eta_{f1}$ and $\eta_{f2}$ are intrinsic parities of $Z_2$ reflections
whose vaules are $+1$ or $-1$, and $\eta_{f0} = \eta_{f1}\eta_{f2}$.
The $\Psi_1$ and $\Psi_2$ are expanded as
\begin{eqnarray}
\hspace{-0.7cm} &~& \Psi_1(x,y)= \sum_{n = -\infty}^{+\infty} \psi_n(x) \exp\left(i\frac{n + \frac{1-\eta_{f0}}{4}}{R} y\right)~,
\label{Psi-1-exp}\\ 
\hspace{-0.7cm} &~& \Psi_2(x,y)= \sum_{n = -\infty}^{+\infty} \eta_{f1} \gamma_5 \psi_n(x) \exp\left(-i\frac{n + \frac{1-\eta_{f0}}{4}}{R} y\right)~,
\label{Psi-2-exp}
\end{eqnarray}
respectively.
Here and hereafter a common normalization factor is omitted.
The extra-coordinate part of kinetic term is given by
\begin{eqnarray}
 i(\overline{\Psi}_1,\overline{\Psi}_2) \Gamma^5
\!\!
\left(
\begin{array}{cc}
\partial_5 +iq_- A_5^{(-)}& 0 \\
0 & \partial_5 -iq_-A_5^{(-)}
\end{array}
\right)
\!\!
\left(
\begin{array}{c}
\Psi_1 \\ 
\Psi_2
\end{array}
\right)~.
\label{Psi-Dy}
\end{eqnarray}
Here and hereafter we omit the SM gauge bosons irrelevant of our discussion
and the gauge coupling of $U(1)^{(-)}$ to avoid a complication.

Upon compactification, the following mass terms appear after integrating on (\ref{Psi-Dy}) over $y$,
\begin{eqnarray}
{\sum_{n = -\infty}^{+\infty}} \frac{n + \frac{1-\eta_{f0}}{4} + q_- \beta}{R} 
\left(\overline{\psi}_{n{\rm L}}(x) \psi_{n{\rm R}}(x) + \overline{\psi}_{n{\rm R}}(x) \psi_{n{\rm L}}(x)\right)~,
\label{psi-mass}
\end{eqnarray}
where $\beta \equiv \langle A_5^{(-)} \rangle R$. 
Note that the field with $\displaystyle{n + \frac{1-\eta_{f0}}{4}} = 0$ can also become massive with $\beta$.
{}From (\ref{psi-mass}), the following one-loop effective potential is obtained
\begin{eqnarray}
V_{\rm eff}^{D}[\beta] = 8C\sum_{n=1}^{\infty}{1 \over n^5}\cos{\left[2\pi n\left(\frac{1-\eta_{f0}}{4} +q_{-}\beta\right)\right]}~,
\label{Psi-Veff}
\end{eqnarray}
where $C\equiv 3/(128\pi^6 R^4)$.
Here and hereafter $\beta$ independent terms are omitted.

For later convenience, we study the fields
defined by $\Psi^{(\pm)} \equiv (\Psi_1 \pm \Psi_2)/\sqrt{2}$.
Using $\Psi^{(\pm)}$, the (\ref{Psi-BC1}) and (\ref{Psi-BC2}) are rewritten as
\begin{eqnarray}
\hspace{0cm} &~& \Psi^{(\pm)}(x,y+2\pi R)=\eta_{f0}\Psi^{(\pm)}(x,y)~,~~
\label{Psipm-BC1}\\ 
\hspace{0cm} &~& \Psi^{(\pm)}(x,-y)=\pm \eta_{f1}\gamma_5\Psi^{(\pm)}(x,y)~,~~
\label{Psipm-BC2}\\
\hspace{0cm} &~& \Psi^{(\pm)}(x,2\pi R-y)=\pm \eta_{f2}\gamma_5\Psi^{(\pm)}(x,y)
\label{Psipm-BC3}
\end{eqnarray}
and (\ref{Psi-Dy}) is rewritten as
\begin{eqnarray}
i(\overline{\Psi}^{(+)},\overline{\Psi}^{(-)})\gamma^5
\!\!
\left(
\begin{array}{cc}
\partial_5  & iq_{-}A_5^{(-)} \\
iq_{-}A_5^{(-)} & \partial_5 
\end{array}
\right)
\!\!
\left(
\begin{array}{c}
\Psi^{(+)} \\ 
\Psi^{(-)}
\end{array}
\right)~.
\label{Psipm-Dy}
\end{eqnarray}
Note that the $\Psi^{(\pm)}$ are the eigenstates of BCs,
but they are not the eigenstates of extra $U(1)$ symmetry.
The non-abelian structure such as $SU(2)$ emerges in (\ref{Psipm-Dy}).

We consider a pair of symplectic-Majorana fermions $\Psi_1$ and $\Psi_2$
whose $U(1)^{(-)}$ charge is $q_-$ and $-q_-$, respectively.
They satisfy the relation such that $\Psi^i = \varepsilon^{ij} C \gamma_5 \overline{\Psi}^{jT}$
where $\varepsilon^{ij} = i \sigma^2$ and $C$ is the charge conjugation matrix.
Majorana fermions acquire the mass $\displaystyle{\left|n + \frac{1-\eta_{f0}}{4} + q_- \beta\right|/R}$ 
upon compactification
and the following one-loop effective potential is obtained 
\begin{eqnarray}
V_{\rm eff}^{M}[\beta] = 4C\sum_{n=1}^{\infty}{1 \over n^5}\cos{\left[2\pi n\left(\frac{1-\eta_{f0}}{4} + q_- \beta\right)\right]}~.
\label{MPsi-Veff}
\end{eqnarray}

\subsubsection{Complex scalar fields}

We consider a pair of complex scalar fields $\Phi_1$ and $\Phi_2$ whose $U(1)^{(-)}$ charge
is $q_-$ and $-q_-$, respectively.
The following BCs are compatible with that (\ref{Psi-L}) should be a single-valued function,
\begin{eqnarray}
\hspace{-0.7cm} &~& \Phi_k(x,y+2\pi R)=\eta_{b0}\Phi_k(x,y)~~~(k=1, 2)~,
\label{Phi-BC1}\\ 
\hspace{-0.7cm} &~& \Phi_1(x,-y)=\eta_{b1}\Phi_2(x,y)~,~~
\Phi_1(x,2\pi R-y)=\eta_{b2}\Phi_2(x,y)~,
\label{Phi-BC2}
\end{eqnarray}
where $\eta_{b1}$ and $\eta_{b2}$ are intrinsic $Z_2$ parities
whose vaules are $+1$ or $-1$, and $\eta_{b0} = \eta_{b1}\eta_{b2}$.
The $\Phi_1$ and $\Phi_2$ are expanded as
\begin{eqnarray}
\hspace{-0.7cm} &~& \Phi_1(x,y)= \sum_{n = -\infty}^{+\infty} \phi_n(x) \exp\left(i\frac{n + \frac{1-\eta_{b0}}{4}}{R} y\right)~,
\label{Phi-1-exp}\\ 
\hspace{-0.7cm} &~& \Phi_2(x,y)= \sum_{n = -\infty}^{+\infty} \eta_{b1} \phi_n(x) \exp\left(-i\frac{n + \frac{1-\eta_{b0}}{4}}{R} y\right)~,
\label{Phi-2-exp}
\end{eqnarray}
respectively.
The extra-coordinate part of kinetic term is given by
\begin{eqnarray}
\left|\left(
\begin{array}{cc}
\partial_5 +iq_-A_5^{(-)}& 0 \\
0 & \partial_5 -iq_-A_5^{(-)}
\end{array}
\right)
\!\!
\left(
\begin{array}{c}
\Phi_1 \\ 
\Phi_2
\end{array}
\right)\right|^2~.
\label{Phi-Dy}
\end{eqnarray}
Upon compactification, the following mass terms appear
\begin{eqnarray}
{\sum_{n = -\infty}^{+\infty}} \left(\frac{n + \frac{1-\eta_{b0}}{4} + q_- \beta}{R}\right)^2
\left|\phi_n(x)\right|^2~.
\label{phi-mass}
\end{eqnarray}
Hence the following one-loop effective potential is induced 
\begin{eqnarray}
V_{\rm eff}^{S}[\beta] = -4C\sum_{n=1}^{\infty}{1 \over n^5}\cos{\left[2\pi n\left(\frac{1-\eta_{b0}}{4} +q_{-}\beta\right)\right]}~.
\label{Psi-Veff}
\end{eqnarray}

The fields defined by $\Phi^{(\pm)} \equiv (\Phi_1 \pm \Phi_2)/\sqrt{2}$
are the eigenstates of BCs
as seen from that the (\ref{Phi-BC1}) and (\ref{Phi-BC2}) are rewritten as
\begin{eqnarray}
\hspace{0cm} &~& \Phi^{(\pm)}(x,y+2\pi R)=\eta_{b0}\Phi^{(\pm)}(x,y)~,~~
\label{Phipm-BC1}\\ 
\hspace{0cm} &~& \Phi^{(\pm)}(x,-y)= \pm \eta_{b1}\Phi^{(\pm)}(x,y)~,~~
\label{Phipm-BC2}\\
\hspace{0cm} &~& \Phi^{(\pm)}(x,2\pi R-y)=\pm \eta_{b2}\Phi^{(\pm)}(x,y)~.
\label{Phipm-BC3}
\end{eqnarray}
On the other hand, the $\Phi^{(\pm)}$ are not the eigenstates of extra $U(1)$ symmetry
as seen from that the (\ref{Phi-Dy}) is rewritten as
\begin{eqnarray}
\left|
\left(
\begin{array}{cc}
\partial_5  & iq_{-}A_5^{(-)} \\
iq_{-}A_5^{(-)} & \partial_5 
\end{array}
\right)
\!\!
\left(
\begin{array}{c}
\Phi^{(+)} \\ 
\Phi^{(-)}
\end{array}
\right)\right|^2~.
\label{Phipm-Dy}
\end{eqnarray}

\section{Our model}

We present an explicit model to realize our scenario.
Our basic principles for BCs are that BCs should be compatible with the single-valued behavior of the Lagrangian density
and preserve $N=1$ SUSY on the brane at $y=0$.
First let us prepare bulk fields whose massless modes at the tree level contain the MSSM particles
and treat them as the eigenstates of BCs.
The minimal sets are given as follows.

~~\\
(i) The member of would-be MSSM gauge multiplets are $(A_M, \Sigma; \lambda^1, \lambda^2)$
whose BCs are given by
\begin{eqnarray}
\hspace{-1.2cm} &~& A_{M}(x,y+2\pi R)= A_{M}(x,y)~,~~ \Sigma(x,y+2\pi R)= \Sigma(x,y)~,~~ 
\nonumber \\
\hspace{-1.2cm} &~& \lambda^{i}(x,y+2\pi R)= \lambda^{i}(x,y)~,
\label{Gauge-BC1}\\ 
\hspace{-1.2cm} &~& A_{\mu}(x,-y)= A_{\mu}(x,y)~,~~ 
\left(
\begin{array}{c}
A_{5} \\ 
\Sigma
\end{array}
\right)(x,-y)
= -\left(
\begin{array}{c}
A_{5} \\ 
\Sigma
\end{array}
\right)(x,y)~,~~ 
\nonumber \\
\hspace{-1.2cm} &~& \lambda^{1}(x,-y)= -\gamma_5 \lambda^{1}(x,y)~,~~ \lambda^{2}(x,-y)= \gamma_5 \lambda^{2}(x,y)~,
\label{Gauge-BC2}\\
\hspace{-1.2cm} &~& A_{\mu}(x,2\pi R-y)= A_{\mu}(x,y)~,~~ 
\left(
\begin{array}{c}
A_{5} \\ 
\Sigma
\end{array}
\right)(x,2\pi R-y)
= -\left(
\begin{array}{c}
A_{5} \\ 
\Sigma
\end{array}
\right)(x,y)~,~~ 
\nonumber \\
\hspace{-1.2cm} &~& \lambda^{1}(x,2\pi R-y)= -\gamma_5\lambda^{1}(x,y)~,~~ \lambda^{2}(x,2\pi R-y)= \gamma_5\lambda^{2}(x,y)~,
\label{Gauge-BC3}
\end{eqnarray}
where $A_M$ is the five-dimensional SM gauge bosons, $\Sigma$ is a real scalar field and $(\lambda^1, \lambda^2)$
are gauginos represented by symplectic-Majorana fermions.
The index indicating the SM gauge group or generators is suppressed.
The massless fields come from $A_{\mu}$ and $\lambda^1_{L}$.
The gauginos can acquire the $\beta$-dependent masses
when $\lambda^1$ and $\lambda^2$ are regarded as $\Psi^{(+)}$ and $\Psi^{(-)}$, respectively,
with $\eta_{1f} = \eta_{2f} = -1$ and yield to the term (\ref{Psipm-Dy}) with a non-vanishing $q_-$.

~~\\
(ii) The member of would-be MSSM matter multiplets are $(\psi^i; \phi^i, \phi^{ci\dagger})$
whose BCs are given by
\begin{eqnarray}
\hspace{-1.4cm} &~& \psi^i(x,y+2\pi R)= \psi^i(x,y)~,
\left(
\begin{array}{c}
\phi^i \\ 
\phi^{ci\dagger}
\end{array}
\right)\!\!(x,y+2\pi R)= 
\left(
\begin{array}{c}
\phi^i \\ 
\phi^{ci\dagger}
\end{array}
\right)\!\!(x,y)~,~ 
\label{Matter-BC1}\\ 
\hspace{-1.4cm} &~& \psi^i(x, -y)= -\gamma_5 \psi^i(x,y)~,
\left(
\begin{array}{c}
\phi^i \\ 
\phi^{ci\dagger}
\end{array}
\right)\!\!(x, -y)= 
\left(
\begin{array}{c}
\phi^i \\ 
-\phi^{ci\dagger}
\end{array}
\right)\!\!(x,y)~,
\label{Matter-BC2}\\
\hspace{-1.4cm} &~& \psi^i(x, 2\pi R-y)= -\gamma_5 \psi^i(x,y)~,
\left(
\begin{array}{c}
\phi^i \\ 
\phi^{ci\dagger}
\end{array}
\right)\!\!(x, 2\pi R-y)= 
\left(
\begin{array}{c}
\phi^i \\ 
-\phi^{ci\dagger}
\end{array}
\right)\!\!(x,y)~,
\label{Matter-BC3}
\end{eqnarray}
where $\psi^i$ are fermions represented by four-component spinors and $(\phi^i, \phi^{ci\dagger})$ are complex scalar fields.
The index $i$ represents particle species.
The massless fields come from $\psi^i_L$ and $\phi^i$
which are the chiral fermions (quarks and leptons) and the corresponding scalar bosons (squarks and sleptons), respectively.
The scalar bosons acquire the $\beta$-dependent masses
when $\phi^i$ and $\phi^{ci\dagger}$ are regarded as $\Phi^{(+)}$ and $\Phi^{(-)}$, respectively,
with $\eta_{1b} = \eta_{2b} = 1$ 
and yield to the term (\ref{Phipm-Dy}) with a non-vanishing $q_-$.

~~\\
(iii) The member of would-be MSSM Higgs multiplets are $(\tilde{h}; h,h^{c\dagger})$ and $(\tilde{\bar{h}}; \bar{h},\bar{h}^{c\dagger})$.
The MSSM Higgsinos come from the fermions $\tilde{h}$ and $\tilde{\bar{h}}$.
The MSSM Higgs bosons stem from the complex scalar fields $(h,h^{c})$ and $(\bar{h},\bar{h}^{c})$.
For simplicity, we impose the same type of BCs as (\ref{Matter-BC1}) -- (\ref{Matter-BC3}) on them
by identifying $\{h, \bar{h}\}$ and $\{h^{c\dagger}, \bar{h}^{c\dagger} \}$ as $\Phi^{(+)}$ and $\Phi^{(-)}$, respectively.
~~\\

The above BCs  (\ref{Gauge-BC1}) -- (\ref{Matter-BC3}) preserve the $N=1$ SUSY on our brane.
We impose the $R$ parity on the model.
Then the theory on our brane is described by the SUSY Lagrangian of the MSSM, using the above particle contents.
The $\mu$ term is assumed to be forbidden at the tree level by some symmetry
because there is no reason to let the magnitude of $\mu$ to be $O(1)$TeV.
On the other hand, the physics in the bulk is depicted by the gauge invariant kinetic terms.

We introduce the abelian gauge boson of $U(1)^{(-)}$
whose BCs are given by (\ref{Apm-BC1}) -- (\ref{Apm-BC3}).\footnote{
The would-be superpartner of $U(1)^{(-)}$ gauge boson is also introduced.
Although the massless mode remains at the low-energy scale after compactification,
it cannot be detected directly because it does not interact with the MSSM fields
in our assignment of gauge quantum numbers.
}
Let us assign the $\lq\lq U(1)^{(-)}$ quantum number" ($q_-$) for the would-be MSSM fields as shown in Table 1.\footnote{
We use the terminology $\lq\lq U(1)^{(-)}$ quantum number" throughout this paper, although it might be unsuitable 
because $\Psi^{(+)}$ ($\Phi^{(+)}$) and $\Psi^{(-)}$ ($\Phi^{(-)}$) are not eigenstates of $U(1)^{(-)}$ gauge symmetry.
Note that the $U(1)^{(-)}$ gauge boson interacts with $\Psi^{(+)}$ ($\Phi^{(+)}$) and $\Psi^{(-)}$ ($\Phi^{(-)}$)
as understood from (\ref{Psipm-Dy}) for fermions and (\ref{Phipm-Dy}) for scalar fields.
We refer to $q_-$ appearing in (\ref{Psipm-Dy}) or (\ref{Phipm-Dy}) as $U(1)^{(-)}$ quantum number.
}
\begin{table}[htb]
\caption[T1]{$U(1)^{(-)}$ quantum numbers}
\begin{center}
\begin{tabular}{c|ccc}
  & $(\lambda^1, \lambda^2)$ & $(\phi^i, \phi^{ci\dagger})$ & $\{h, \bar{h}, h^{c\dagger}, \bar{h}^{c\dagger} \}$ \\ \hline
$q_-$ & $q_{\lambda}$  & $q_{\phi^i}$ & $q_{h}$ \\ 
\end{tabular}
\end{center}
\end{table}
We assume that the $U(1)^{(-)}$ quantum number for other fields are zero.
Then the following one-loop effective potential is induced 
\begin{eqnarray}
&~& V_{\rm eff}^{\mbox{\tiny{MSSM}}}[\beta] 
= -4C\sum_{i} \sum_{n=1}^{\infty}{1 \over n^5}\cos{\left[2\pi n q_{\phi^i}\beta\right]}
\nonumber \\
&~& ~~~~  + 4C \sum_{n=1}^{\infty}{1 \over n^5}\cos{\left[2\pi n q_{\lambda}\beta\right]} 
-8C \sum_{n=1}^{\infty}{1 \over n^5}\cos{\left[2\pi n q_{h}\beta\right]} ~.
\label{MSSM-Veff}
\end{eqnarray}
If the magnitude of $q_{\phi^i}$ and/or $q_h$ is the same order of $q_{\lambda}$,
the minimum of $V_{\rm eff}^{\mbox{\tiny{MSSM}}}[\beta]$ is given by $\beta = 0$
and then soft SUSY breaking terms are not induced.
It is possible to obtain the minimum with $q_{\phi^i}\beta = 1/2$ by changing BCs with $\eta_{b0} = -1$,
but some fields can acquire heavy masses of $O(1/R)$ comparable to those of Kaluza-Klein modes
and the MSSM with soft SUSY breaking parameters of $O(1)$TeV and a small extra dimension much less than 
$O(10^{-18})$m cannot be derived.

Here we list problems including the above one.
\begin{itemize}
\item[(1)] How can we obtain the non-vanishing VEV of $A_5^{(-)}$ ?

\item[(2)] Can the magnitude of soft SUSY breaking masses be $O(1)$TeV ?

\item[(3)] How can we obtain the $\mu$ term ?
\end{itemize}

We introduce extra fields to solve the first problem.
A simple one is a pair of SM singlet complex scalar fields $\Phi^{(\pm)}$ whose BCs are given by
(\ref{Phipm-BC1}) -- (\ref{Phipm-BC3}) with $\eta_{b0} = - 1$.\footnote{
A fermion with $Q_- = 0$ is assumed to be present to keep the $N=1$ SUSY on the brane.
}
Let the $U(1)^{(-)}$ quantum number of $\Phi^{(+)}$ and $\Phi^{(-)}$ be $q_{\Phi}$.
The following one-loop effective potential is induced 
\begin{eqnarray}
V_{\rm eff}^{\Phi}[\beta] 
 = -4C\sum_{i} \sum_{n=1}^{\infty}{1 \over n^5}\cos{\left[2\pi n \left(\frac{1}{2} + q_{\Phi}\beta\right)\right]}
\label{Psi-Veff}
\end{eqnarray}
and the minimum of $V_{\rm eff}^{\Phi}[\beta]$ is given by $q_{\Phi} \beta = 1/2$.
Then the first problem is solved if the contribution of $V_{\rm eff}^{\Phi}[\beta]$ dominates over that of 
$V_{\rm eff}^{\mbox{\tiny{MSSM}}}[\beta]$ in the determination for the minimum of the potential.
The intriguing possibility is that the magnitude of $q_{\Phi}$ is much bigger than that of $q_{\lambda}$,
$q_{\phi^i}$ and $q_{h}$.

Let the magnitude of $q_{\Phi}$ be $O(1)$ and we tackle the second problem.
In this case, the magnitude of $\langle A_5^{(-)} \rangle$ is estimated as $O(1/R)$
and that of  soft SUSY breaking masses is estimated as $O(q_{\rm sp}/R)$ 
where $q_{\lambda}$, $q_{\phi^i}$ and $q_{h}$ are denoted as $q_{\rm sp}$ as a whole.
If $1/R$ is $O(10^{16})$GeV, a tiny charge such that $q_{\rm sp} =O(10^{-13})$ is required
to obtain masses of $O(1)$TeV.
We find that
{\it a large hierarchy of $q_{\rm sp}/q_{\Phi} = O(R/\mbox{TeV}^{-1})$ is necessary 
in order to obtain the soft SUSY breaking masses of $O(1)$TeV.}
It is a difficult problem whether such a tiny charge or a large charge hierarchy is derived naturally.
This is one of problems in our scenario.

Finally we consider the last problem.
We introduce another abelian gauge symmetry denoted as $U(1)'^{(-)}$
and extend the BCs for the member of would-be MSSM Higgs multiplets.
We regard $\tilde{h}$ and $\tilde{\bar{h}}$ as $\Psi^{(+)}$ and $\Psi^{(-)}$, respectively
and impose the same type of BCs as (\ref{Psipm-BC1}) -- (\ref{Psipm-BC3}) on them with $\eta_{f1} = \eta_{f2} = -1$.
Then the massless fields in $\tilde{h}$ and $\tilde{\bar{h}}$ obtain the masses such as 
$q'_{\tilde{h}} \langle {A'}_5^{(-)} \rangle$ from the term (\ref{Psipm-Dy})
where $q'_{\tilde{h}}$ and ${A'}_5^{(-)}$ are the $\lq\lq U(1)'^{(-)}$ quantum number" of $\tilde{h}$ and $\tilde{\bar{h}}$
and the $U(1)'^{(-)}$ gauge boson, respectively.

Before studying the Higgs bosons, we consider a quartet of complex scalar fields 
$\Phi_1$, $\Phi_2$, $\Phi_3$ and $\Phi_4$ whose $U(1)^{(-)} \times U(1)'^{(-)}$
quantum numbers are given by $(q_-, q'_-)$, $(-q_-, -q'_-)$, $(-q_-, q'_-)$
and $(q_-, -q'_-)$.
The BCs of ($\Phi_1$, $\Phi_2$) are given by (\ref{Phi-BC1}) -- (\ref{Phi-BC2}) with $\eta_{b1} = \eta_{b2} = 1$
and those of ($\Phi_3$, $\Phi_4$) are the same.
Let us compose the eigenstates of BCs as
\begin{eqnarray}
\hspace{-1.4cm}&~& \Phi^{(+)} = c(\Phi_1 + \Phi_2) + s(\Phi_3 + \Phi_4)~,~~
\Phi'^{(+)} = -s(\Phi_1 + \Phi_2) +c(\Phi_3 + \Phi_4)~,
\label{Phi++} \\
\hspace{-1.4cm}&~& \Phi^{(-)} = c(\Phi_1 - \Phi_2) + s(\Phi_3 - \Phi_4)~,~~
\Phi'^{(-)} = -s(\Phi_1 - \Phi_2) + c(\Phi_3 - \Phi_4)~,
\label{Phi--}
\end{eqnarray}
up to a normalization factor.
Here $c$ and $s$ are defined as $c \equiv \cos\theta$ and $s \equiv \sin\theta$, using a mixing angle $\theta$.
The $(\Phi^{(+)}, \Phi'^{(+)})$ and $(\Phi^{(-)}, \Phi'^{(-)})$ 
satisfy the same type of BCs as (\ref{Phipm-BC1}) -- (\ref{Phipm-BC3}) with $\eta_{b1} = \eta_{b2} = 1$.
Using them, 
the extra-coordinate part of kinetic term is given by
\begin{eqnarray}
\left|
\left(
\begin{array}{cccc}
\partial_5 & 0 & N' & N \\
0 & \partial_5 & N & N' \\
N' & N & \partial_5 & 0 \\
N & N' & 0 & \partial_5 
\end{array}
\right)
\left(
\begin{array}{c}
\Phi^{(+)} \\ 
\Phi'^{(+)} \\
\Phi^{(-)} \\ 
\Phi'^{(-)}
\end{array}
\right)\right|^2~,
\label{Phipmpm-Dy}
\end{eqnarray}
where $N$ and $N'$ are defined by
\begin{eqnarray}
N \equiv 2i sc q_{-}A_5^{(-)}~,~~
N' \equiv i (c^2-s^2) q_{-}A_5^{(-)} + iq'_{-}{A'}_5^{(-)}~.
\label{N}
\end{eqnarray}
The fields with $n=0$ appear in $\Phi^{(+)}$ and $\Phi'^{(+)}$ 
and the mass-squared matrix of them is given by
\begin{eqnarray}
\left(
\begin{array}{cc}
m_{\tiny\mbox{D}}^2 & m_{\tiny\mbox{O}}^2\\
m_{\tiny\mbox{O}}^2 & m_{\tiny\mbox{D}}^2
\end{array}
\right)~,
\label{Phi++mass}
\end{eqnarray}
where $m_{\tiny\mbox{D}}^2$ and $m_{\tiny\mbox{O}}^2$ are 
\begin{eqnarray}
&~& m_{\tiny\mbox{D}}^2 = 4 s^2 c^2 \left(q_{-}\langle A_5^{(-)} \rangle \right)^2 
 + \left[(c^2-s^2)q_{-}\langle A_5^{(-)} \rangle + q'_{-}\langle {A'}_5^{(-)}\rangle\right]^2~,
\label{mD}\\
&~& m_{\tiny\mbox{O}}^2 = 4sc q_{-}\langle A_5^{(-)} \rangle 
\left[(c^2-s^2)q_{-}\langle A_5^{(-)} \rangle + q'_{-}\langle {A'}_5^{(-)}\rangle\right]~,
\label{Phi++mass}
\end{eqnarray}
respectively.
Note that the above mentioned BCs respect the $N=1$ SUSY on the brane 
when $(\Phi^{(\pm)}, \Phi'^{(\pm)})$ and $\Psi^{(\pm)}$ make up two sets of hypermultiplets.

{}From the above observation, the fields with $n=0$ in $\Phi^{(+)}$ and $\Phi'^{(+)}$ 
can be identified by two Higgs doublets $h_u$ and $h_d^{\dagger}$, respectively.
For simplicity, we choose $\theta = \pi/4$.
Then we obtain the mass-squared matrix such that
\begin{eqnarray}
\left(
\begin{array}{cc}
\left(q_{-}\langle A_5^{(-)} \rangle \right)^2 + \left(q'_{-}\langle {A'}_5^{(-)}\rangle\right)^2 
& 2q_{-}q'_{-}\langle A_5^{(-)} \rangle\langle {A'}_5^{(-)}\rangle \\
2q_{-}q'_{-}\langle A_5^{(-)} \rangle\langle {A'}_5^{(-)}\rangle & 
\left(q_{-}\langle A_5^{(-)} \rangle \right)^2 + \left(q'_{-}\langle {A'}_5^{(-)}\rangle\right)^2
\end{array}
\right)~.
\label{Phi++mass2}
\end{eqnarray}
In this case, $\left(q_{-}\langle A_5^{(-)} \rangle \right)^2$ is the soft SUSY breaking scalar mass,
$q'_{-}\langle {A'}_5^{(-)}\rangle$ is the $\mu$ parameter if $q'_{-} = q'_{\tilde{h}}$,
and 
$2q_{-}q'_{-}\langle A_5^{(-)} \rangle\langle {A'}_5^{(-)}\rangle$ is the $B$ parameter ($B\mu$).

The remainig problems are how the non-vanishing $\gamma \equiv \langle {A'}_5^{(-)}\rangle R$ can be obtained
and whether the magnitude of $\mu$ can be $O(1)$TeV.
We introduce a pair of Dirac fermions $\Psi^{(\pm)}$ with $\eta_{f1} = \eta_{f2} = 1$.\footnote{
Two pairs of complex scalar fields with vanishing $U(1)'^{(-)}$ charge are assumed to be present to keep the $N=1$ SUSY on the brane.
}
Let the $U(1)'^{(-)}$ quantum number of $\Psi^{(+)}$ and $\Psi^{(-)}$ be $q'_{\Psi}$.
Then the following one-loop effective potential is induced 
\begin{eqnarray}
V_{\rm eff}^{\Psi}[\gamma] 
= 8C\sum_{i} \sum_{n=1}^{\infty}{1 \over n^5}\cos{\left[2\pi n q'_{\Psi}\gamma\right]}~.
\label{Psi-Veff}
\end{eqnarray}
The minimum of $V_{\rm eff}^{\Psi}[\gamma]$ is given by $q'_{\Psi}\gamma = 1/2$.
If the magnitude of $q'_{\Psi}$ is much bigger than that of $q_{h}$ and $q'_{\tilde{h}}$,
the contribution of $V_{\rm eff}^{\Psi}[\gamma]$ dominates over others in the determination of $\gamma$
and then $\langle {A'}_5^{(-)}\rangle$
stabilizes as $q'_{\Psi}\gamma = 1/2$.

In the case with $q_{\Phi} \beta =1/2$, $q'_{\Psi}\gamma = 1/2$ and $\theta = \pi/4$, the $\mu$ and $B$ parameters are given by
\begin{eqnarray}
\mu = q'_{\tilde{h}} \langle {A'}_5^{(-)}\rangle = \frac{1}{2R} \frac{q'_{\tilde{h}}}{q'_{\Psi}}~,~~
B\mu = q_h q'_{\tilde{h}} \langle A_5^{(-)} \rangle \langle {A'}_5^{(-)}\rangle = \frac{1}{4R^2} \frac{q_h}{q_{\Phi}} \frac{q'_{\tilde{h}}}{q'_{\Psi}}~,
\label{mu-B}
\end{eqnarray}
respectively.
If we require that $\mu = O(1)$TeV and $B\mu = O(1)$TeV$^2$, 
{\it we need a large hierarchy of $q'_{\tilde{h}}/q'_{\Psi} = O(R/\mbox{TeV}^{-1})$ and $q_{h}/q_{\Phi} = O(R/\mbox{TeV}^{-1})$.}

In this way, the following soft SUSY breaking mass terms and $\mu$ term are derived,
\begin{eqnarray}
&~& \mathcal{L}_{\tiny{\mbox{soft}}} = -\left(\frac{1}{2} \sum_{a} M_{\lambda}^a {\lambda}^a \lambda^a + \mbox{h.c.}\right)
- \sum_i m_i^2 |\phi^i|^2 
\nonumber \\
&~& ~~~~~~~~~~ - m_{h_u}^2 |h_u|^2 - m_{h_d}^2 |h_d|^2 - B\mu\left( h_u h_d + \mbox{h.c.}\right)~,
\label{Lsoft}\\
&~& \mathcal{L}_{\mu} = -\mu^2\left(|h_u|^2 + |h_d|^2\right) - \left(\mu {\tilde{h}}_u \tilde{h}_d+ \mbox{h.c.}\right)~,
\label{Lmu}
\end{eqnarray}
where $M_{\lambda^a} = q_{\lambda^a} \langle {A}_5^{(-)}\rangle$,
$\displaystyle{m_i^2 = \left(q_{\phi^i} \langle {A}_5^{(-)}\rangle\right)^2}$,
$\displaystyle{m_{h_u}^2 = m_{h_d}^2 = \left(q_{h} \langle {A}_5^{(-)}\rangle\right)^2}$,
$B \mu = q_h q'_{\tilde{h}} \langle A_5^{(-)} \rangle \langle {A'}_5^{(-)}\rangle$
and $\mu = q'_{\tilde{h}} \langle {A'}_5^{(-)}\rangle$ for $\theta = \pi/4$.
Here $\lambda^a$ are the MSSM gauginos whose $U(1)^{(-)}$ quantum number is assigned by $q_{\lambda^a}$,
$\phi^i$ are the MSSM sfermions
and $(h_u, h_d)$ and $(\tilde{h}_u, \tilde{h}_d)$ are the MSSM Higgs bosons and Higgsinos.
We find that the breakdown of electroweak symmetry can occur radiatively using the same analysis in \cite{BH&N1}.

Our model does not suffer from the SUSY CP problem because all soft SUSY breaking masses and $\mu$ parameter
are real-valued.
The flavor changing neutral current (FCNC) processes can be suppressed enough
if the $U(1)^{(-)}$ quantum number of the relevant sfermions has a common value.

Our scenario can be applied to the SUSY $SU(5)$ orbifold GUT.\cite{K2,H&N}
By the introduction of the following matrices $(P_0, P_1)$ relating $Z_2$ reflections under $y=0$ and $y=\pi R$
to the BCs of two sets of Higgs multiplets,
\begin{eqnarray}
P_0 = \mbox{diag}(1,1,1,1,1)~,~~ P_1 = \mbox{diag}(-1,-1,-1,1,1)~,
\label{P0P1}
\end{eqnarray}
the triplet-doublet splittings are realized elegantly.

\section{Conclusions and discussion}

We have proposed the mechanism that the soft SUSY breaking masses
and $\mu$ parameter can be induced from the dynamical rearrangement of the local symmetries
such as $U(1)^{(-)} \times U(1)'^{(-)}$ in a five-dimensional model.
It offers to a solution of $\mu$ problem if there is a large hierarchy among the relevant $U(1)$ charge of Higgsinos 
and the SM singlets which stabilize the extra-dimensional coordinate of $U(1)$ gauge boson.

Our proposal is regarded as a loophole from the no go theorem that 
{\it soft SUSY breaking parameters and $\mu$ parameter of $O(1)$TeV cannot be obtained via the Hosotani mechanism
from any SUSY QFT without SUSY breaking sources and with flat small extra dimensions.}
We have relaxed the assumption that the SUSY in starting QFT is partially broken by introducing exotic $U(1)$ gauge symmetries.
It is possible to generate mass parameters with approriate size by choosing $U(1)$ charges with a suitable magnitude.
Furthermore it is possible to build the non-abelian structure such as $SU(2)$
by using the difference between the eigenstates of $U(1)$ gauge symmetries and those of BCs.

The big problem in our scenario is the origin of exotic $U(1)$ symmetries.
The exoitics come from the following three features.
Different quantum numbers are assigned for bosons and fermions in (would-be) SUSY multiplets.
The eigenstates of gauge symmetries are mixtures of chiral SUSY multiplets
and anti-chiral ones, though the eigenstates of BCs are chiral SUSY multiplets and anti-chiral ones in themselves.
The magnitude of exotic $U(1)$ charges is extremely small of $O(R/\mbox{TeV}^{-1})$.
{}From the third feature, we fancy a strange possibility that such $U(1)$ symmetries come from non-abelian gauge groups
and $U(1)$ charged fields are solitons appearing on the breakdown of gauge symmetry.
This idea has stemed from by reference to the Witten effect.\cite{W}
There, however, remain open questions whether a similar effect can occur in a higher-dimensional SUSY theory
and the masslessness of $U(1)$ charged fields is guaranteed by any reason.

\section*{Acknowledgements}
This work was supported in part by scientific grants from the Ministry of Education, Culture,
Sports, Science and Technology under Grant Nos.~22540272 and 21244036 (Y.K.)
and No.~23$\cdot$9368 (T.M.).

\end{document}